\documentclass[a4paper,12pt]{article}

\pdfoutput=1
\usepackage{amsmath}
\usepackage{a4}
\usepackage{pstricks,pst-node,pst-text,pst-3d}
\usepackage{amsfonts}
\usepackage{amssymb}
\usepackage{graphicx,graphics,epsfig}
\usepackage{geometry}
\usepackage[english]{babel}
\usepackage[utf8]{inputenc}
\usepackage[T1]{fontenc}
 \usepackage{float}
\usepackage{hyperref}
\usepackage{ulem}
\usepackage{caption}
\usepackage{cite}
\usepackage{caption}
\usepackage{subfigure}  
\usepackage{mathrsfs}  
\usepackage{bbold}

\def\noi{\noindent}
\def\non{\nonumber}

\newcommand{\cosb}{c_\beta}
\newcommand{\sinb}{s_\beta}
\newcommand{\cosa}{c_\alpha}
\newcommand{\sina}{s_\alpha}
\newcommand{\sinapb}{s_{\alpha+\beta}}
\newcommand{\cosapb}{c_{\alpha+\beta}}
\newcommand{\sinbma}{s_{\beta-\alpha}}
\newcommand{\cosbma}{c_{\beta-\alpha}}
\newcommand{\sbt}{s_{2\beta}}
\newcommand{\cbt}{c_{2\beta}}
\newcommand{\sat}{s_{2\alpha}}
\newcommand{\cat}{c_{2\alpha}}

\newcommand{\msbar}{\overline{\rm MS}}

\def\l{\lambda}

\def\b{\beta}

\def\G{\Gamma}

\def\ba{\begin{array}}
\def\ea{\end{array}}
\def\bea{\begin{eqnarray}}
\def\eea{\end{eqnarray}}

\newcommand{\Asla}{\not{\hbox{\kern-3.5pt $A$}}}
\newcommand{\Gsla}{\not{\hbox{\kern-3.5pt $G$}}}
\newcommand{\Wsla}{\not{\hbox{\kern-3.5pt $W$}}}
\newcommand{\Zsla}{\not{\hbox{\kern-3.5pt $Z$}}}

\newcommand{\Dslash}{\not{\hbox{\kern-4pt $D$}}}
\newcommand{\pslash}{\not{\hbox{\kern-2.3pt $p$}}}

\newcommand{\CUV}{C_{UV}}

\def\lsim{\;\raise0.3ex\hbox{$<$\kern-0.75em\raise-1.1ex\hbox{$\sim$}}\;}
\def\gsim{\;\raise0.3ex\hbox{$>$\kern-0.75em\raise-1.1ex\hbox{$\sim$}}\;}

\def\l{\lambda}

\def\b{\beta}

\def\ba{\begin{array}}
\def\ea{\end{array}}
\def\bea{\begin{eqnarray}}
\def\eea{\end{eqnarray}}

\def\lsim{\;\raise0.3ex\hbox{$<$\kern-0.75em\raise-1.1ex\hbox{$\sim$}}\;}
\def\gsim{\;\raise0.3ex\hbox{$>$\kern-0.75em\raise-1.1ex\hbox{$\sim$}}\;}

\begin{document}

\begin{titlepage}
\begin{center}
\vspace*{-1cm}
\begin{flushright}
LPT-Orsay-17-39\\ 
KCL-PH-TH/2017-42
\end{flushright}

\vspace*{1.6cm}
{\large\bf The neutral Higgs self--couplings in the (h)MSSM} 

\vspace*{1cm}\renewcommand{\thefootnote}{\fnsymbol{footnote}}

{\large 
G.~Chalons$^{1}$\footnote[1]{Email: guillaume.chalons@th.u-psud.fr},
A.~Djouadi$^{1}$\footnote[2]{Email: abdelhak.djouadi@th.u-psud.fr}
and 
J.~Quevillon$^{2}$\footnote[3]{Email: jeremie.quevillon@kcl.ac.uk}.
}

\vspace*{0.3cm}


\renewcommand{\thefootnote}{\arabic{footnote}}

\vspace*{1cm} 
{\normalsize \it 
$^1\,$ Laboratoire de Physique Th\'eorique, CNRS UMR8627 and  
Univiversit\'e Paris-Sud, \\ Universit\'{e} Paris-Saclay, 91045
Orsay, France\\[2mm]
$^2\,$ {King's College London, Strand, London, WC2R 2LS, United Kingdom\\[2mm]
}}

\vspace{0.3cm}

\end{center}
\begin{abstract}

We consider the Minimal Supersymmetric extension of the Standard Model  in the
regime where the supersymmetric breaking scale is extermely large. In  this
MSSM,  not only the Higgs masses will be affected by large  radiative
corrections, the dominant part of which is provided by the third generation
quark/squark sector, but also the various self--couplings among the Higgs
states. In this note, assuming that squarks are extremely heavy, we evaluate
the  next-to-leading order radiative corrections to the  two neutral CP--even
Higgs self--couplings $\lambda_{H  h h}$ and $\lambda_{h h h}$ and to the
partial decay width $\Gamma(H \to h h)$ that are most relevant at the LHC. The
calculation is  performed using an effective field theory approach that resums
the large logarithmic squark contributions and allows to keep under control the
perturbative expansion.  Since the direct loop vertex corrections are generally
missing in this effective approach, we have properly renormalised the effective
theory to take them into account. Finally, we perform a comparison of the
results in this effective MSSM  with those obtained in a much simpler way in the
so--called  hMSSM approach  in which the mass value for the  lightest Higgs
boson $M_{h}=125$ GeV is  used as an input.  We show that the hMSSM  provides a
reasonably good approximation of the corrected self--couplings and $H \to hh$
decay rate and, hence, it can be used also in these cases.

\end{abstract}

\end{titlepage}

\subsection*{1. Introduction}

Dedicated  analyses of the data collected at the LHC have so far shown excellent
agreement between the observed $125$ GeV Higgs boson \cite{observedH} and the
scalar particle that is predicted  in the  Standard  Model (SM) of particle
physics \cite{Higgs}. However, it is widely believed that this  model is simply
an effective theory valid at low energies and that new  physics beyond it should
manifest itself at a scale not too far from the  TeV scale.  This new physics
should be thus probed at the LHC with higher  luminosity and/or at future
collider experiments where one should  either discover direct evidence  of new
particles or detect some deviations  from SM predictions.

In this respect, reconstructing the Higgs potential at the LHC and  eventually
at future high energy colliders is a major undertaking both  on the experimental
and theoretical sides \cite{TripleExp,Triple}. The parameters  involved in the Higgs
potential  feature relations  among the Higgs  masses and their self-couplings
that are crucial to determine in order  to fully understand the nature of the
Higgs particle. Higgs boson pair production probes directly the triple Higgs
self--coupling which, in  the SM, is entirely fixed in terms of the Higgs mass
and the vacuum  expectation value. In models with extended Higgs sectors, the
value of  the self--coupling of the observed Higgs state can not only differ
from  the SM value but, in addition, other Higgs bosons can be exchanged  in the
processes in which this light state is doubly produced. Therefore, measuring a
deviation in the pair production of the SM--like  Higgs boson would point to a
non-minimal Higgs sector and, hence, to   physics beyond the SM.

Supersymmetry (SUSY) is a typical beyond the SM scenario possessing such an
extended Higgs sector. As a matter of fact, in low energy SUSY scenarios, at
least two Higgs doublet fields $H_u$ and $H_d$ are required to break the
electroweak symmetry and to generate masses to the known gauge bosons and
fermions. In its simplest incarnation, the Minimal Supersymmetric Standard 
Model (MSSM), the spectrum consists of five states \cite{MSSM-Higgs}:  two
charged scalars $H^\pm$, a CP-odd $A$ and two CP-even $h,H$ neutral  scalars.
The phenomenology of the Higgs  sector is described entirely by  two input
parameters, one Higgs mass that is usually taken to be that of the pseudoscalar
$A$ boson $M_{A}$ and the ratio $\tan \b$ of the vacuum expectation values of
the two doublet fields, $\tan \b=v_u/v_d$. However, in the $(M_{A},\tan \b)$
parameter space, the prediction for the mass of the lightest (observed) Higgs
boson is at odds with the measured value at the LHC, $M_{h} \approx 125$ GeV,
unless the large radiative corrections from the other SUSY sectors, most notably
from the stop/top sector, are included to raise its mass at the desired value
\cite{CR-leading,accurate-hmass}.  This renders the MSSM parameter space survey
a very complicated task and benchmark scenarios, such as those presented in
Refs.~\cite{Carena:2013ytb,Bagnaschi:2015hka},  were designed to ease
interpretation of the data.

Nevertheless,  in a general MSSM framework, it is  rather difficult to satisfy
the constraint $M_{h} \approx 125$ GeV in all cases and, to circumvent this
shortcoming, a minimal and almost model independent approach, called the hMSSM
\cite{hMSSM,Djouadi:2015jea}, has been put forward. In this framework, by taking
the measured mass value $M_{h} \approx 125$ GeV as an input, one removes the
dependence of the Higgs sector on the dominant radiative corrections and, hence,
on the additional SUSY parameters. The hMSSM has been shown to  provide a very
good description of the MSSM Higgs mass spectra and the mixing angle $\alpha$ in
the CP--even Higgs sector  \cite{Bagnaschi:2015hka,Djouadi:2015jea}. As a bonus,
it  allows us to access the entire $(M_{A},\tan \b)$ parameter space without
being in conflict with the LHC data. In particular the low $\tan \b$ regime can
be probed, at the expense of assuming a very high SUSY scale $M_{\rm SUSY}$ such
that the  radiative corrections (that grow logarithmically with $M_{\rm SUSY}$)
allow the mass  $M_h$ to attain the value of $ 125$ GeV.

In the MSSM, not only the Higgs masses and the mixing angle $\alpha$ are
affected by large radiative corrections, but also the various 
self--couplings between the Higgs states. In the case where the SUSY scale is
extremely large, evaluating   the radiative corrections through fixed-order
perturbative calculations is seriously questionable and using Effective Field
Theory (EFT) methods seems a more appropriate approach; for a recent discussion
see  Ref.~\cite{Lee:2015uza}. Indeed, in case of a large mass hierarchy between
the SUSY and electroweak scales, EFT techniques resum the large logarithmic
contributions and enable to keep under control the perturbative expansions.
However, when such techniques are used to compute some physical processes such
as decays or production rates, they may miss direct vertex ``genuine''
corrections, \textit{e.g} momentum dependent corrections, that are contained in
the fixed-order diagrammatic calculation, which can be significant. This is 
particularly the case of the triple Higgs couplings and, especially,  the decay 
$H \to hh$ that involve the self--coupling $\lambda_{Hhh}$ and in which the 
vertex corrections, e.g. involving top/stop loops, are important
\cite{Bagnaschi:2015hka}. In the hMSSM approach, the effects of the large
logarithmic corrections are captured in the neutral Higgs masses and mixing
angle but, there also, the genuine vertex corrections should be included. 

In this brief note, we combine EFT methods and fixed-order calculations to
derive the next-to-leading order (NLO) corrections  to the two neutral triple
CP--even Higgs couplings $\lambda_{hhh}$ and $\lambda_{Hhh}$ and the rate of the
decay mode $H\to hh$  that are most relevant at the LHC. These are first
evaluated in an effective MSSM obtained from matching the full MSSM to an
effective two--Higgs doublet model (2HDM) below the scale $M_{\rm SUSY}$ (where
the squarks have been integrated out) and that we renormalise to obtain
ultraviolet finite results.  We then estimate the size of the additional loop
(vertex)  corrections and compare with the hMSSM predictions to assess to which
extent the two approaches differ. We show that the hMSSM approach provides a
reasonably good approximation and thus can be used even in the case of the
triple Higgs couplings and the rates for the double production of the $h$ state.

\subsection*{2. The Higgs self--couplings and $\mathbf{H\to hh}$ at NLO  \label{compframework}}

We first require the Higgs boson observed at the LHC to be the lightest Higgs
scalar of the MSSM $h$ with a mass $M_{h}=125$ GeV.  At low $\tan \beta$  and
moderate $M_A$ values, since the SUSY scale is defined to be the  geometric
average of the masses of the two stop partners of the heavy  top quark, $M_{\rm
SUSY} = \sqrt { m_{\tilde t_1} m_{\tilde t_2} }$,  multi-TeV stop squark masses
are required to reach the Higgs mass value above as the leading radiative
correction is of the form $M_t^4 \log(M_t^2/M_{\rm SUSY}^2)$  \cite{CR-leading}
with $M_t$ the top quark mass. Fixed-order NLO calculations performed in this
regime are thus not reliable as they lead to too large radiative corrections. 
In the very large $M_{\rm SUSY}$ regime, EFT methods lead to much more reliable
results  since the large logarithms induced by the multi-TeV stops masses are
resummed and absorbed into effective couplings. The current state-of-the-art
calculations regarding the radiative corrections to the MSSM Higgs boson masses
and the mixing angle $\alpha$  combine fixed-order calculations and EFT
techniques \cite{accurate-hmass}. 

The calculation of the  radiative corrections to the production and decay rates
of the lightest Higgs particle in the MSSM are, in turn, less precise.  This is
particularly true for the important decay modes $H\to h h$ and  $(gg\to) \,
h^{*} \to h h$  that appear in double  $h$ production and, thus, probe the
CP--even neutral Higgs self--couplings.  In this note, we will concentrate on
the NLO radiative corrections to the on-shell partial decay width  $\Gamma(H \to
h h)$ and to the two Higgs self--couplings $\lambda_{Hhh}$ and $\lambda_{hhh}$
that are most relevant for LHC physics. We will work in an effective field
theory obtained from matching the full MSSM to an effective two--Higgs doublet
model (2HDM) below the scale $M_{\rm SUSY}$, where the squarks have been
integrated out. The matching is performed using the public code \texttt{MhEFT}
and details about the procedure and the code in itself can be found in Ref.~\cite{Lee:2015uza}.
Nevertheless, for simplicity and as a first step, we do  not fully match the
MSSM
to a general 2HDM and we restrict ourselves to  the following scalar potential,
\begin{eqnarray} 
\label{VhMSSM2} \mathscr{V} &=& m_{H_d}^2 |H_d|^2 +  m_{H_u}^2 |H_u|^2+ m_{12}^2
(H_u \cdot H_d + h.c)  + \frac{\l_1}{2}| H_d|^4 + \frac{\l_2}{2}|H_u|^4 \non \\ 
& & + \l_3 |H_d|^2 |H_u|^2 +\l_4|H_u \cdot H_d|^2
\end{eqnarray}\noi  
where $v$ is the ``true'' vacuum expectation value (vev), $v =\sqrt{
v_d^2+v_u^2}\simeq 174\,\mbox{GeV}$. The SU(2) doublets $H_d,H_u$ with
hypercharge $Y = \mp 1$, respectively, are given by 
\begin{equation} H_d = \begin{pmatrix} v_d + (\phi_d
- i \varphi_d)/\sqrt{2} \\ -\phi_d^- \end{pmatrix},~ H_u = \begin{pmatrix}
\phi_u^+ \\ v_u + (\phi_u + i \varphi_u )/\sqrt{2} \end{pmatrix},
\end{equation}\noi 

In terms of the gauge boson masses which we will use as inputs, a  tree-level
matching with the MSSM would simply lead to the relations
\begin{equation} 
\l_1 = \l_2 = \frac{M_Z^2}{2 v^2}, \quad \l_3 = \frac{2 M_W^2 -
M_Z^2}{2 v^2}, \quad \l_4 = -\frac{M_W^2}{v^2} 
\end{equation}\noi

Compared to the general 2HDM we have restricted ourselves only to $\l_1, \cdots,
\l_4$ quartic couplings but the model is still renormalisable. The reason behind
this choice is that at tree-level in the MSSM, SUSY imposes that the 
$\l_5,\l_6,\l_7$ quartic couplings are absent\footnote{As well as the bilinear
terms  $m^2_{H_{d/u}}$ and $m_{12}^2$ which are only generated once SUSY
is (softly) broken and essential for electroweak symmetry breaking.}. However it
may well be that models of supersymmetry breaking can provide a direct
contribution to these parameters but technically these contributions would not
correspond to a soft-SUSY breaking mechanism, as is assumed in the MSSM. Of
course, even in such a scenario these additional operators are generated
through renormalisation group running but, being absent from the tree level
potential, they give rise to finite subleading corrections and, in a first
approximation, they can be neglected. In addition, we do not take into account
possible higher-dimensional operators that could also modify the Higgs
properties. For Higgs phenomenology, the multi-TeV stops masses affect mostly
the renormalisation of the $|H_u|^4$ operator and thus can be absorbed in a
redefinition of its Wilson coefficient/coupling (the so-called ``$\l_2$''
parameter) using EFT techniques. Matching the full MSSM to an effective 2HDM
will thus absorb the largest radiative corrections and NLO calculations within
this theory will, a priori, lead to much more stable results. 

Since we will be only interested in computing the by far dominant radiative
corrections induced by third generation quarks/squarks to the Higgs decays 
modes that we are considering, the Yukawa sector of our EFT consists of a type
II 2HDM; see Ref.~\cite{Branco:2011iw} for example. In fact, the Yukawa
Lagrangian below $M_{\rm SUSY}$ should be written as the most general
Higgs-fermion Yukawa couplings, and not a type II-like Lagrangian. In
particular, at large $\tan \b$, it is well-known that wrong Higgs coupling, 
namely couplings of $H_u$ to down type fermions, are generated and can be
significant. However, for the regime of $\tan \b$ that we are interested in ($1
\leq \tan \b \lesssim 10$ as will be seen later), these should not be
numerically large and we can neglect them. The trilinear Higgs couplings that
will be relevant for our analysis are the following,
\begin{align}
\label{Gh3}
\l_{hhh} &= -3\sqrt{2}\left(v_d \sina^3 \l_1 -
\cosa\left(v_u \cosa^2 \l_2 + \sina (-v_d \cosa + v_u \sina
(\l_3+\l_4\right)\right)  \\
 \label{GHhh}
\l_{Hhh} &= \frac{3 v_d \sina \sat }{\sqrt{2}}\l_1 + 3\sqrt{2}v_u \cosa^2
\sina \l_2 + \frac{v_d \cosa (1+ 3(\cosa^2 - 3 \sina^2))+ v_u \sina
(1 - 3 (2 \cosa^2 + \cat ))}{2 \sqrt{2}}\left(\l_3+\l_4\right) \\
\label{GHHh}
\l_{H H h} &= - 3 \sqrt{2} v_d \cosa^2 \sina \l_1 + \frac{3 v_u \sina
\sat \l_2}{\sqrt{2}} +  \frac{v_u \cosa (1+ 3(\cosa^2 - 3 \sina^2))- v_d \sina
(1 - 3 (2 \cosa^2 + \cat ))}{2 \sqrt{2}}\left(\l_3+\l_4\right)
\end{align}\noi 
where we use the abbreviations $s_\alpha = \sin \alpha$ etc.. with $\alpha$
the mixing angle in the CP--even Higgs sector given by 
\begin{equation}
 \tan 2 \alpha = \frac{\sbt (2 v^2 (\l_3+\l_4)) -
M_{A}^2}{2v^2(\l_1 \cosb^2-\sinb^2 \l_2)-M_{A}^2 \cbt}\quad
\mbox{and}\quad\tan \b = v_u/v_d
\end{equation}\noi 

The only inputs that we need are the four parameters $\l_1, \cdots, \l_4$, $\tan
\b$ and the pseudo--scalar mass $M_{A}$. We obtain these parameters from the
code \texttt{MhEFT} after running them down to the scale of the pseudscalar mass
$M_{A}$. More precisely we set as input values $\l_1,\cdots, \l_4 \equiv
\l_1(M_A), \cdots, \l_4(M_{A})$ and $\tan \b \equiv \tan \b(M_{A})$. As regards
the renormalisation of the model, all fields and parameters introduced so far
are considered as bare parameters. Shifts are then introduced for the Lagrangian
parameters and the fields with the notation that a bare quantity is labeled as
$X_0$. All bare quantities ($X_0$) are then decomposed into renormalised ($X$)
and counterterms ($\delta X$) quantities as $X_0 \to X + \delta X$. The
counterterms to $\l_1, \cdots, \l_4$ and $\tan \b$ will be defined in the
$\msbar$ scheme. The divergent parts $\delta \l_1^{\msbar}, \cdots,
\delta \l_4^{\msbar}$ can be obtained from the beta functions given, for
example, in the Appendix of
\cite{Chowdhury:2015yja}, retaining only the top and bottom Yukawa
contributions. The $\tan \b$ counterterm is defined by
\begin{equation}
 \frac{\delta t_\b^{\msbar}}{t_\b} = \frac{3}{32 \pi^2}\left(  Y_b^2 -  Y_t^2
\right) \CUV
\end{equation}\noi
with $\CUV = 1/\epsilon - \gamma_E + \ln(4\pi)$ and $Y_f$ the Yukawa coupling
of the corresponding fermion $f$. The shifts on the doublet vevs $v_d$ and $v_d$
are related to the counterterm of $\tan \beta$ through,
\begin{equation}
 (v_d)_0 \to v_d \left( 1 - \sinb^2 \frac{\delta t_\b^{\msbar}}{t_\b}\right),
\quad (v_u)_0 \to v_u \left( 1 +\cosb^2 \frac{\delta t_\b^{\msbar}}{t_\b}\right)
\end{equation}\noi

The pseudoscalar Higgs mass is renormalised on-shell and we do not introduce a
counterterm for the angle $\alpha$ since we already consider it as a
renormalised quantity, see Ref.~\cite{Baro:2008bg} for more details. As
explained in the latter reference, the renormalisation of the mixing between
$H$ and $h$ still has to be performed and is transferred into a counterterm
to the off-diagonal entry of the CP--even Higgs mass matrix, that we denote as
$\delta  M^2_{hH}$. This counterterm can be obtained from the following
equation,
\begin{eqnarray}
 \delta  M^2_{h H} &=& v^2 \sat \left(\sinb^2 \delta \l_2^{\msbar} - \cosb^2
\delta \l_1^{\msbar}\right) + v^2 \cat \sbt \left(\delta \l_3^{\msbar} +\delta
\l_4^{\msbar} \right) \non \\
& &+ 2
v^2 \left( \sat \left(\cosb^2 \l_2 - \sinb^2 \l_1\right) + \cat \sbt \left(\l_3
+ \l_4\right) \right)\frac{\delta v}{v} \non \\
& & + \left(4 v^2 \left( 2\sat \sbt \left(\l_1 + \l_2 \right) +  
\cat \left(\l_3 +\l_4\right)\right) - \frac{M^2_{A}}{2}\cat 
\right)\sbt \frac{\delta t_\b^{\msbar}}{t_\b} \non \\
& & + \cosbma \left(\cosbma^2 - 3 \sinbma^2 + 3 
\right)\frac{\delta T_{h}}{4\sqrt{2} v} + \sinbma \left(\sinbma^2 -3
\cosbma^2 +  3 \right)\frac{\delta T_{H}}{4\sqrt{2} v} \\
& & - \cosbma \sinbma \delta M^2_{A}
\end{eqnarray}\noi 
To fully define this counterterm we need to determine the additional
counterterms $\delta v, \delta T_{h}, \delta T_{H}$ and we refer to
Ref.~\cite{Baro:2008bg} for their explicit definition. To completely
remove the divergencies arising in the one-loop computation of the
Higgs-to-Higgs decays, the three field renormalisation constants $\delta
Z_{h},\delta Z_{H}$ and $\delta Z_{hH/Hh}$ are needed and we again 
follow the on-shell prescription used in Ref.~\cite{Baro:2008bg} to define
them. We have now all the necessary ingredients to compute the one-loop finite
corrections to the trilinear couplings we are interested in. 

The $H \to h h$ partial decay rate is then given by
\begin{equation}
 \G(H \to h h) = \frac{|{\cal A}(H \to h h)|^2}{32 \pi
M_{H}}\sqrt{1-\frac{4 M_{h}^2}{M^2_{H}}}
\end{equation}\noi 
where the amplitude $\cal A$ is the sum of the tree level plus one loop
amplitudes ${\cal A} = {\cal A}_0 + {\cal A}_1$. The amplitude ${\cal A}_0$ is
simply given by the coupling $\l_{Hhh}$ in eq.~(\ref{GHhh}). The one-loop amplitude ${\cal A}_1$ corresponds to,
\begin{eqnarray}
{\cal A}(H \to h h) &=& \left(1 + \delta Z_{h} + \frac{1}{2}\delta
Z_{H}+\frac{\delta \l_{Hhh}}{\l_{Hhh}}\right)\l_{Hhh} +
\frac{1}{2}\delta Z_{hH}
\l_{hhh}+ \delta Z_{Hh}\l_{HHh}\non \\
& & + {\Lambda}_{Hhh}^{(1)}(M_{h}^2,M_{h}^2,M^2_{H})
\end{eqnarray}\noi 
where ${\Lambda}_{Hhh}^{(1)}$ is the unrenormalised
one-loop proper vertex. $\delta \l_{Hhh}$ is the counterterm to the
coupling given by eq.~(\ref{GHhh}), obtained after performing the shifts on the
input parameters defining it. The two other trilinear Higgs couplings were
given in eqs.~(\ref{Gh3}),(\ref{GHHh}).

We define the finite one-loop correction to the triple $h$ self--coupling by,
\begin{eqnarray}
 \l_{hhh}^{(1)}(M_{h}^2,M_{h}^2,4 M^2_{h}) &=& \left( 1+
\frac{3}{2}\delta Z_{h}+\frac{\delta
\l_{hhh}}{\l_{hhh}}\right)\l_{hhh}  +
\frac{3}{2} \delta Z_{H h} \l_{H h h}\non \\ & & +
{\Lambda}_{h hh}^{(1)}(M_{h}^2,M_{h}^2,4 M^2_{h})
\end{eqnarray}\noi 
where the first line is for the counterterm contribution and the second line 
for the
unrenormalised proper vertex correction. Again, the counterterm $\delta
\l_{hhh}$ is obtained after performing the appropriate shifts on
eq.~(\ref{Gh3}).

To compute these one-loop observables our renormalisation program has been
implemented in the \texttt{SloopS}
code \cite{Baro:2008bg,Baro:2009gn}, to perform our numerical
investigation, to which we now turn.

\subsection*{3. Numerical analysis}

To perform our numerical analysis, we have implemented the Lagrangian defined by
eq.~(\ref{VhMSSM2}) within \texttt{SloopS}, a code for the automated generation
and evaluation of any cross section for any model. \texttt{SloopS}  is an
interface between two packages, the \texttt{LanHEP} program \cite{LanHEP}, where
the Lagrangian of the model is defined as well as the one-loop shifts on the
parameters and the bundle \texttt{FormCalc/FeynArts/LoopTools} \cite{FFL}  that
we will call \texttt{FFL} for short. \texttt{LanHEP} automatically derives the
Feynman rules of the model, including the counterterm contribution. The
generated model files are then passed to the \texttt{FFL} bundle which takes
care of computing at the one-loop level the observables. Some studies have
already been performed with \texttt{SloopS} in Higgs phenomenology in the MSSM
\cite{Baro:2008bg} and recently in the next-to-MSSM (NMSSM)
\cite{Belanger:2017rgu}.

Our procedure to compute the partial decay width and the one-loop correction to
the triple $h$ couplings at NLO within the effective field theory that we
defined in eq.~(\ref{VhMSSM2}) is the following. First, we perform a scan over
the parameters $\tan \b$ and $M_{A}$ with the code \texttt{MhEFT} in order to
obtain a mass for the lightest MSSM Higgs boson $M_{h}= 125\pm 5\,\mbox{GeV}$,
where 5 GeV is assumed to be a very gross estimate of the theoretical
uncertainty in the determitation of this mass in the MSSM. These two parameters
are scanned in the following ranges, $ \tan \beta \in [1;10]$ and $M_{A} \in
[240;600]$ GeV. These intervals represent a regime where the decay $H \to h h$
is phenomenologically relevant. Indeed, far from the $t\bar{t}$ threshold, $M_{H} \gg 2M_{t}$, the decay mode $H \to t \bar t$ becomes largely
dominant and all the other modes are then irrelevant. 

With the help of the \texttt{MhEFT} code, we have performed two scans with
different SUSY common mass scales, $M_{\rm SUSY} = 10$ and $50$ TeV and with
all sfermion soft masses assumed to be equal to this value. The trilinear soft
SUSY-breaking terms are set to $A_{b/\tau} = 5$ TeV while the trilinear soft
SUSY-breaking term $A_t$ is defined through the stop mixing parameter  $X_t =
A_t - \mu \cot \b$ and is set to $X_t/M_{\rm SUSY} = \sqrt{6}$,  corresponding
to the so-called maximal mixing scenario. The higgsino mass parameter $\mu$  and
the gaugino soft mass terms $M_1$ and $M_2$ which play a minor role  have been
set to  $\mu = M_1 =M_2 = 2$ TeV. The parameters $\l_1, \cdots, \l_4$ are then
extracted at the scale $M_{A}$ and fed into \texttt{SloopS} for the computation
of the NLO correction to the trilinear Higgs couplings. The top Yukawa coupling
is defined from the top pole mass which is set to $M_t = 172.5$ GeV and
similarly for the bottom Yukawa coupling with $M_b = 4.62$ GeV. Since we work in
an effective theory where all heavy sfermions have been integrated out, their
loop contributions (in particular the $\tilde t$ and $\tilde b$  ones) are
already encoded into the parameters $\l_1, \cdots, \l_4$. Thus, we only included
the top and bottom contribution into the relevant loops for computing the NLO
corrections as all other contributions are sub-dominant.

\begin{figure}[htbp]
 \begin{center}
  \includegraphics[width=.32\textwidth]{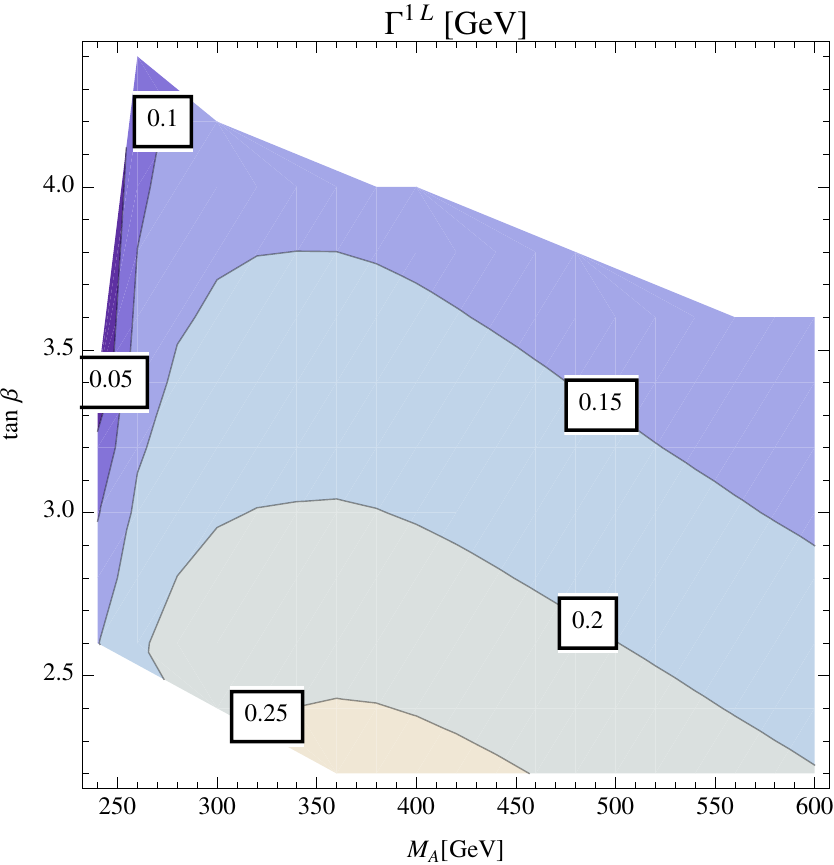}
  \includegraphics[width=.32\textwidth]{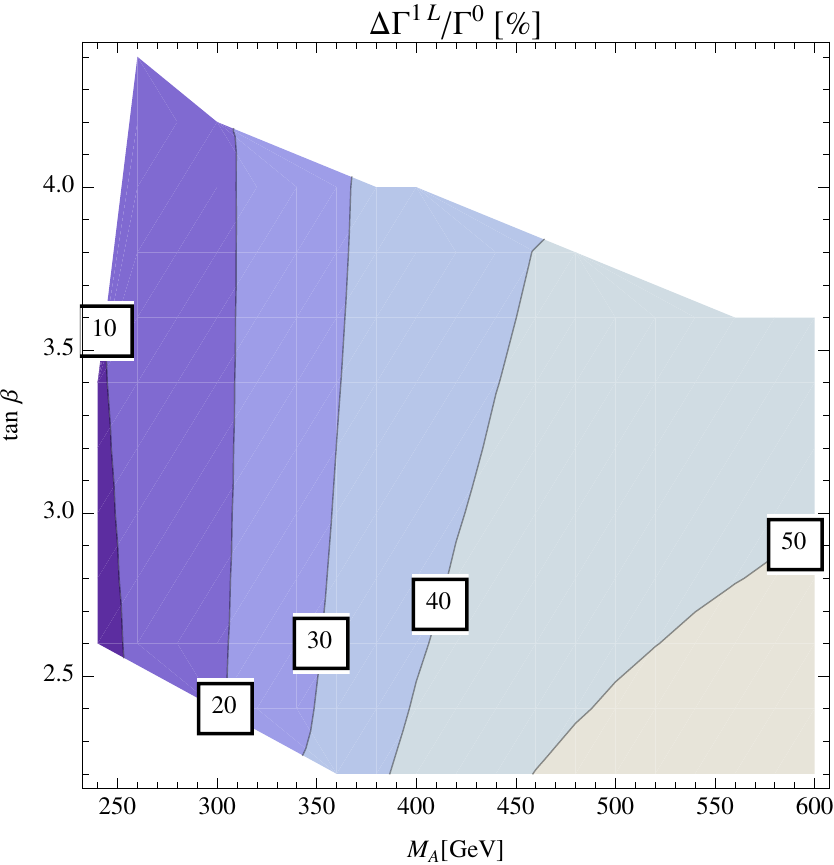}
  \includegraphics[width=.32\textwidth]{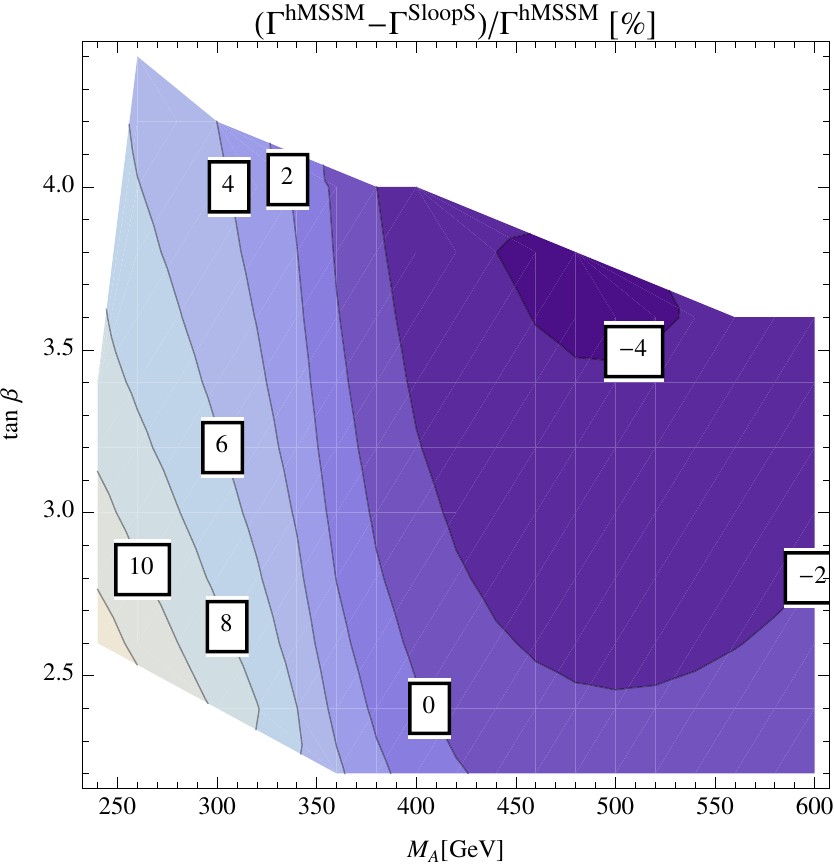}
 \end{center}\caption{\label{fig:XtMS50}  Left: Decay
width (in GeV) $\Gamma(H \to h h)$ computed at NLO in the $(M_{A},t_\b)$
plane. Center: Relative size of the one-loop corrections to $\Gamma(H \to h
h)$ in percent. Right: Relative difference in $\Gamma(H \to h h)$
between the predictions from the hMSSM approach in eq.~(\ref{GHhhhMSSM}) and our
computational procedure. In all three cases, we set $M_{\rm SUSY}=50$ TeV.}
\end{figure}

\begin{figure}[htbp]
 \begin{center}
  \includegraphics[width=.32\textwidth]{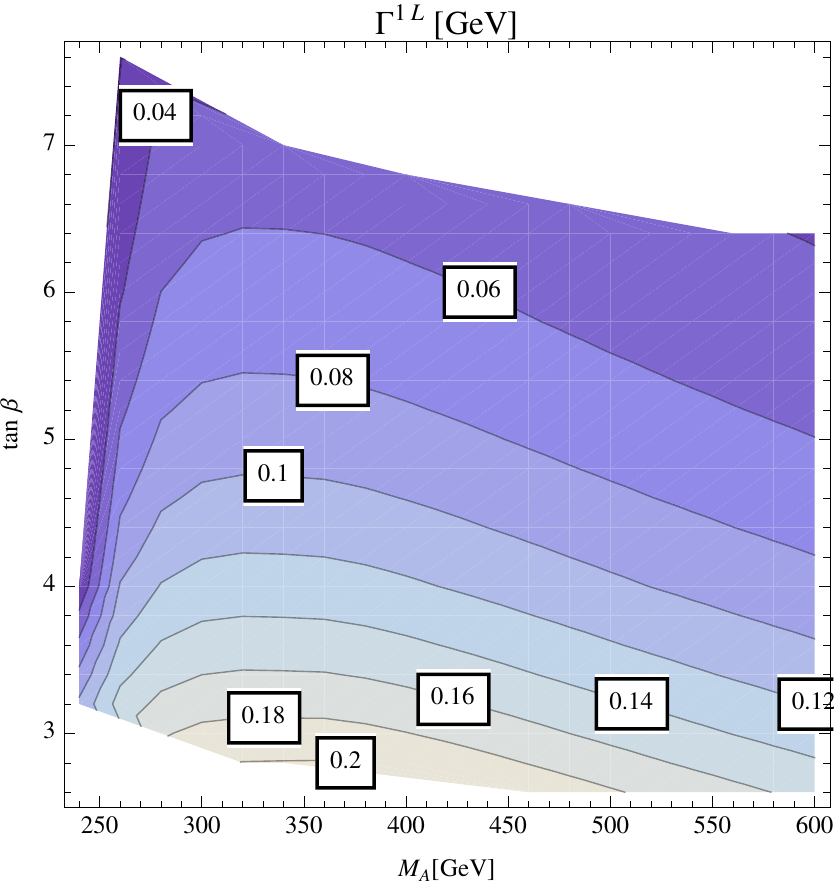}
  \includegraphics[width=.32\textwidth]{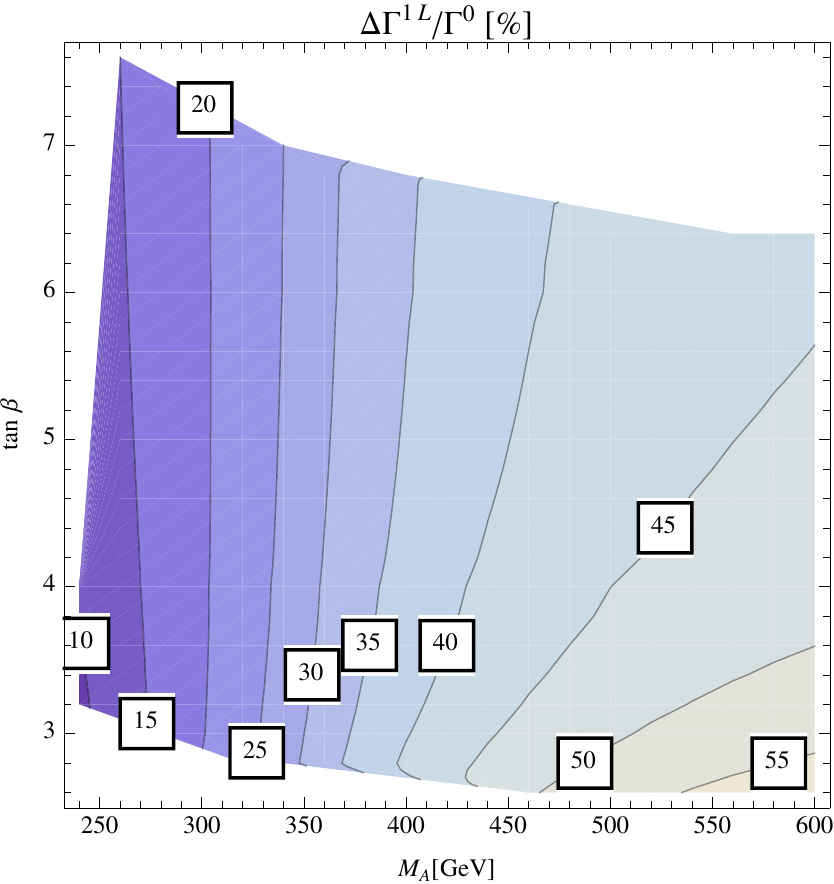}
  \includegraphics[width=.32\textwidth]{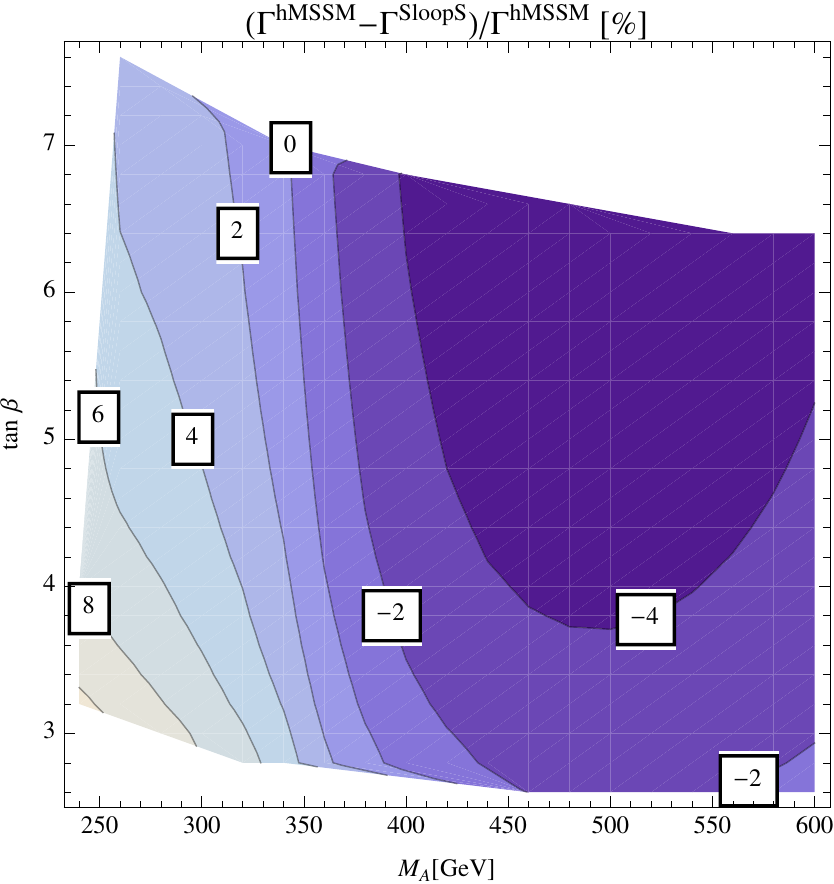}
 \end{center}\caption{\label{fig:XtMS10} Same as in Fig.\ref{fig:XtMS50}
except that $M_{\rm SUSY}=10$ TeV.}
\end{figure}

\begin{figure}[htbp]
 \begin{center}
  \includegraphics[width=.38\textwidth]{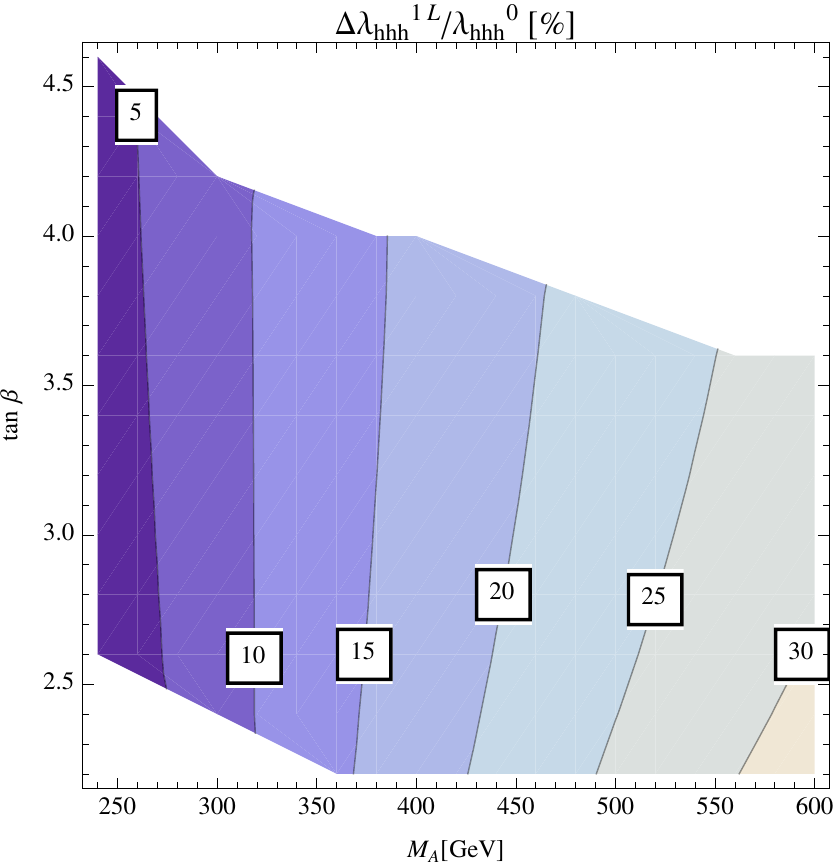}~~~
  \includegraphics[width=.38\textwidth]{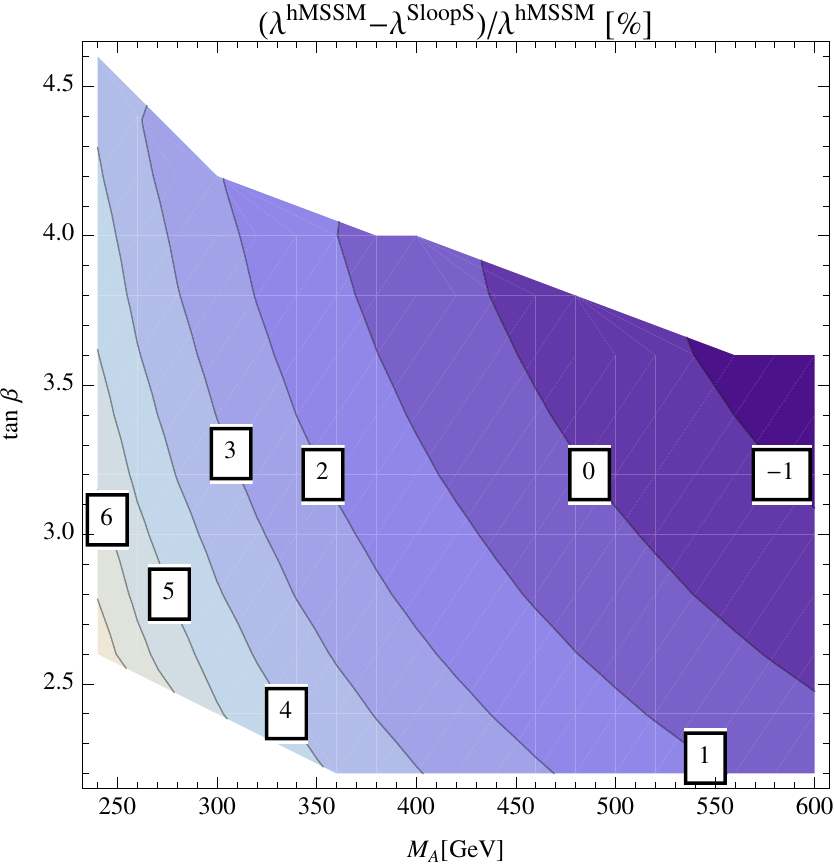}
 \end{center}\caption{\label{fig:hhhXtMS50} Left: Relative
size of the one-loop corrections to $\l_{hhh}$ in percent. Right: Relative
difference in $\l_{hhh}$ between the predictions from the hMSSM approach in eq.~(\ref{G3hMSSM}) and our computational procedure. In both cases, we set $M_{\rm SUSY}=50$ TeV.}
\end{figure}

\begin{figure}[htbp]
 \begin{center}
  \includegraphics[width=.38\textwidth]{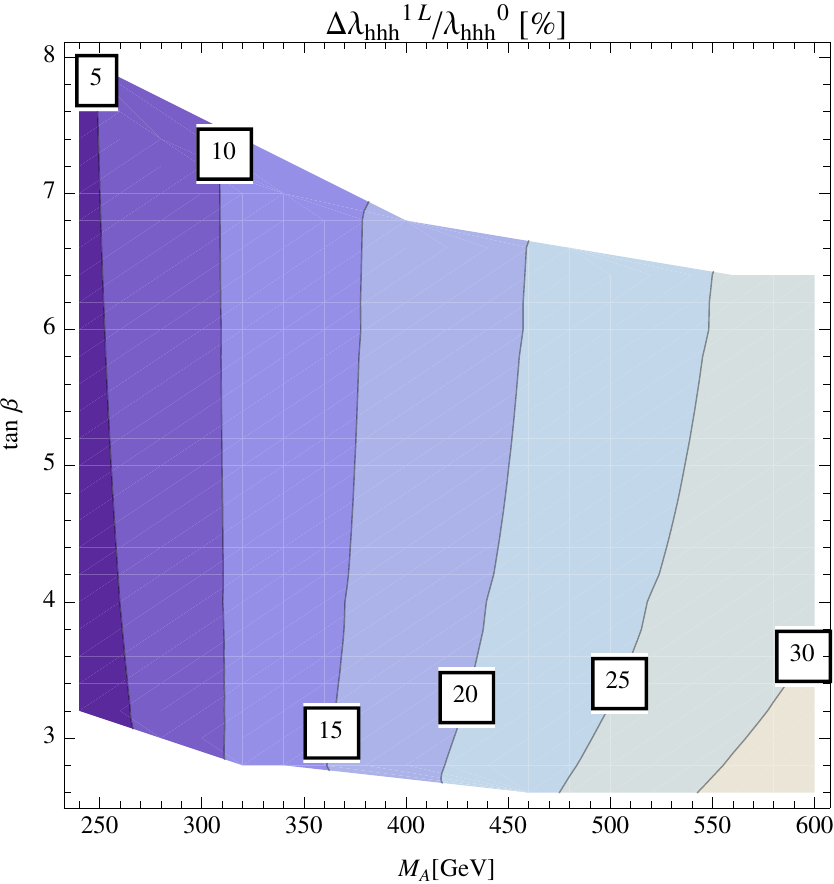}~~~
  \includegraphics[width=.38\textwidth]{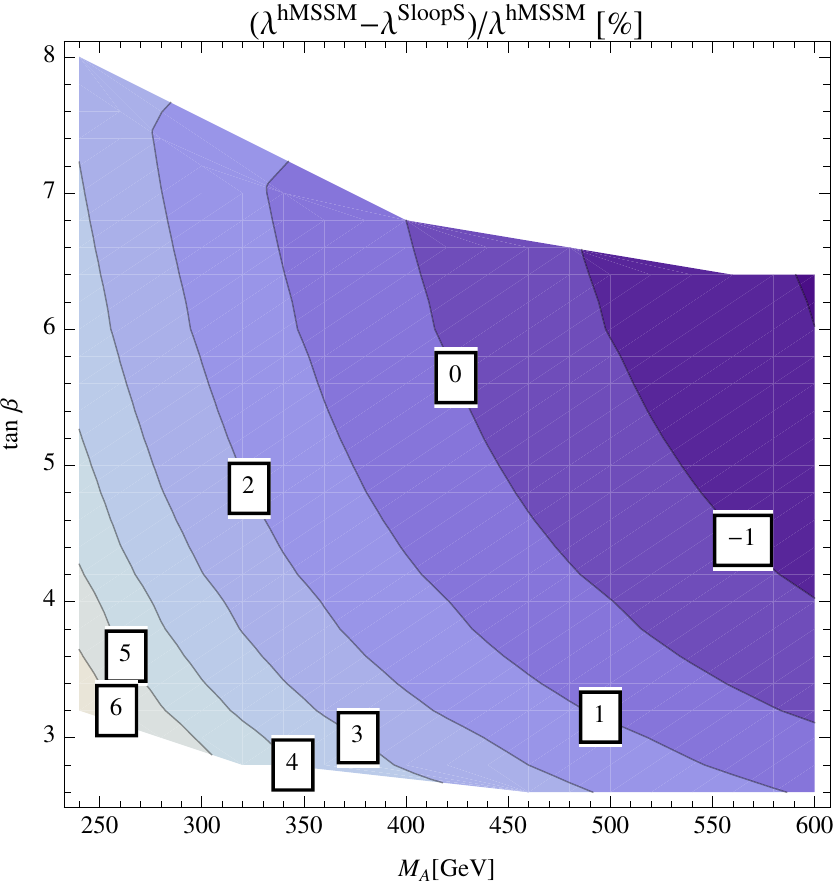}
 \end{center}\caption{\label{fig:hhhXtMS10}Same as in
Fig.\ref{fig:hhhXtMS50} except that $M_{\rm SUSY}=10$ TeV.}
\end{figure}

Our results are displayed in Figures \ref{fig:XtMS50} and
\ref{fig:XtMS10} for the $H \to h h$ partial width and Figures
\ref{fig:hhhXtMS50} and \ref{fig:hhhXtMS10} for the NLO correction to the
triple $h$ coupling.

The left panels in Figs. \ref{fig:XtMS50} and \ref{fig:XtMS10} represent the
total NLO decay width $\Gamma(H \to h h)$ obtained with our computational
framework. The centeral panels are for the relative one-loop corrections to the
tree--level decay widths and in the  right panels, we perform a comparison with
the prediction for $\Gamma(H \to h h)$ derived from the hMSSM approach
\cite{hMSSM,Djouadi:2015jea}. In this approach the trilinear coupling of the
heavy scalar to two light ones reads, in units of $M_Z^2/2v^2$ (in passing we
also give the expression for the triple light Higgs coupling to be discussed
below),
\begin{align}
\label{GHhhhMSSM}
 \l_{Hhh} &= 2 \sat \sinapb - \cat \cosapb + 3 \frac{\Delta {\cal
M}_{22}^2 }{M_Z^2} \frac{\sina}{\sinb} \cosa^2 \\
\label{G3hMSSM}
 \l_{hhh} &= 3 \cat \sinapb + 3 \frac{\Delta {\cal
M}_{22}^2 }{M_Z^2} \frac{\cosa}{\sinb} \cosa^2
\end{align}\noi 
where $\Delta {\cal M}_{22}^2$ is obtained from the known value of $M_{h}$:
\begin{equation}
 \Delta {\cal M}_{22}^2 = \frac{M_{h}^2 (M_{A}^2 + M_{Z}^2 -
M_{h}^2)- M_{A}^2 M_{Z}^2 \cbt}{M_{Z}^2 \cosb^2 + M_{A}^2 \sinb^2 -
M_{h}^2}
\end{equation}\noi

Both Figures \ref{fig:XtMS50} and \ref{fig:XtMS10} are limited from below and
above by the constraint on the mass $M_{h}$ and exhibit qualitatively the  same
features. In Fig.\ref{fig:XtMS10}, we see that lowering $M_{\rm SUSY}$ down to
10 TeV allows for a larger $\tan \b$ range than in Fig.\ref{fig:XtMS50}  in
which $M_{\rm SUSY}=50$ TeV. Nevertheless this enables us to probe smaller
values of $\tan \b$.

Let us first comment on the total decay width $\Gamma(H \to h h)$ which is
displayed in the left panels of Fig.\ref{fig:XtMS50} and \ref{fig:XtMS10}. In
both cases we observe that $\Gamma(H \to h h)$ is favored for low values of
$\tan \b$ and $M_{A}$ and decreases more steeply with respect to increasing
values of $\tan \beta$ than with respect to $M_{A}$. The panels in the center of
Fig.\ref{fig:XtMS50} and \ref{fig:XtMS10} show that our perturbative calculation
is relatively under control,  with the NLO corrections reaching their maximum
for low values of $\tan \b$ and higher values for $M_{A}$. Although the
radiative corrections are substantial, they are much more reasonable than if we
had computed $\Gamma(H \to h h)$ in the plain MSSM. Indeed, in this case, they
can reach several hundreds of percent, jeopardizing the validity of a pure
fixed-order calculation in this regime of low $\tan \b$. In the right panels
where we performed a comparison between our approach and the hMSSM one, we can
see that both predictions agree well and the hMSSM indeed captures the bulk of
the radiative corrections in this regime of low $\tan \b$ values and moderate
$M_{A}$ value for the $H \to h h$ decay. 

We next turn to the discussion of the one-loop corrections to the triple $h$
coupling, where our results are displayed in Fig.\ref{fig:hhhXtMS50} and
Fig.\ref{fig:hhhXtMS10}. We only present the relative one-loop corrections on
the left panel and in the right panel we perform a comparison with the
prediction from the hMSSM obtained from eq.~(\ref{G3hMSSM}).

The results for this coupling exhibit the same features as for the $H \to h
h$ decay rate. Here also, the effective approach allows  to keep the
radiative correction under control in the region where the heavy Higgs decay is
phenomenologically interesting, i.e. low $\tan \beta$ and $M_H \lesssim 350$
GeV. For both decays we observe that the size of the loop corrections grow
mainly with the pseudoscalar mass $M_{A}$ and are almost independent of $\tan
\b$. Again, the main visible difference between Fig.\ref{fig:hhhXtMS50} and
Fig.\ref{fig:hhhXtMS10} is that for the former lower values can be probed
because of the light Higgs mass constraint. The right panels of
Fig.\ref{fig:hhhXtMS50} and Fig.\ref{fig:hhhXtMS10} display the comparison
between the hMSSM prediction and the one obtained from the procedure  detailed
previously. In this case also, the hMSSM prediction is very close to the more
complete calculation performed here. 

\subsection*{4. Conclusion}

In this note, we have considered the neutral Higgs boson self-couplings  $\l_{h
h h}$ and $\l_{H h h}$ and performed a comparison of their predicted  values in
both the hMSSM and an effective MSSM approaches,  in a regime where the  SUSY
scale $M_{\rm SUSY}$ is extremely large. The use of an effective MSSM theory
instead of the full MSSM is necessary at very high $M_{\rm SUSY}$ as large
logarithmic corrections involving this scale and corresponding to the squark
masses appear and need to be resummed in order to obtain reliable results. To
this purpose, we have matched the full  MSSM to an effective theory which
corresponds to a restricted (but still rather general) renormalisable
two-Higgs-doublet model of type II. To include also the genuine direct or vertex
corrections due to the third generation quarks, which are absent in a
renormalisation  group improved calculation,  we have renormalised the effective
theory in the $\msbar$ scheme. 

We have then shown that the NLO radiative corrections are well under control but
they can be substantial for large $M_{A}$ values. Nevertheless for the partial
width of the process   $H \to h h$, the corrections are small  in the regions
where the decay  is phenomenologically relevant. The comparison with the hMSSM
predictions for the neutral self-couplings revealed that this simple approach 
still provides a reasonably good approximation (the deviations are smaller than
$10\%$ and are even less in general) and, hence, the hMSSM approach can be used
not only to determine the MSSM Higgs masses and the mixing angle  $\alpha$, as
shown in previous studies, but also to evaluate these Higgs self-couplings.  

In a future work,  we plan  to refine our analysis by including in the effective
MSSM theory the subleading  contributions of the gauge and scalar bosons and  of
the gauginos and higgsinos. We will also improve our predictions by matching the
full MSSM to a general 2HDM, thereby including also the subleading contributions
coming from the $\l_{5,6,7}$ Lagrangian parameters and their renormalisation. 
In addition, an extension of the analysis to the other Higgs self-couplings, 
in particular those involving the pseudoscalar and charged Higgs states, is
foreseen.

\subsubsection*{Acknowledgements}
G.C. and A.D. are supported by the ERC advanced grant Higgs@LHC. G.C. would
like to thank the LAPTh in Annecy where large parts of this work were realised.
The work of J.Q is supported by the STFC Grant ST/L000326/1. We acknowledge
useful discussions with Gabriel Lee and Carlos Wagner.\vspace*{-2mm}



\end{document}